\documentclass[10.5pt,a4paper,oneside]{extarticle}
\usepackage[T1]{fontenc} 
\usepackage[left=25mm,right=20.9mm,top=4.25cm,bottom=3.75cm,headsep=0.25cm,footskip=0.25cm]{geometry}

\usepackage[english]{babel}
    \usepackage[affil-it]{authblk}
    \usepackage{etoolbox}
    \usepackage{lmodern}
    \usepackage{setspace}
    \usepackage{amssymb,amsthm,amsmath}
    \usepackage{tikz}
    \usepackage{listings}
    \usepackage{color}
    \usepackage{caption}
    \usepackage{subcaption}
    \usepackage{placeins}
    \usepackage{booktabs}
    \usepackage{enumerate}
    \usepackage{graphicx}
    \usepackage{titlesec}
    \usepackage{mathptmx}
    \usepackage{anyfontsize}
    \usepackage{t1enc}
    \usepackage[utf8]{inputenc}
    \usepackage{fancyhdr}
    \usepackage{float,xspace}
    \usepackage{scalefnt}
	\usepackage{comment}
	\usepackage[para]{footmisc}
	
\usepackage{soul} 
\usepackage{tabulary}
\usepackage[colorlinks=true,linkcolor=blue]{hyperref} 

\usepackage{textcase}

\usepackage[colorinlistoftodos]{todonotes}

\titlespacing*{\section}
{0pt}{12pt}{0pt}
\titlespacing*{\subsection}
{0pt}{12pt}{0pt}
\titlespacing*{\subsubsection}
{0pt}{12pt}{0pt}
\titleformat{\section}
  {\huge\sffamily\Large\bfseries\color{black}}
  {\thesection}{1em}{}
    \titleformat{\subsection}[block]{\bfseries}{\thesubsection}{.5em}{}
    \titleformat{\subsubsection}[block]{\bfseries}{\thesubsubsection}{.5em}{}

\titleformat{\section}{\fontsize{12}{19}\bfseries}{\thesection}{1em}{}

\usepackage{mathptmx} 

\makeatletter

\patchcmd{\@maketitle}{\LARGE \@title}{\fontsize{14}{19.2}\selectfont\@title}{}{} 
\makeatother
\title
{
	\vspace{-5cm}
	\begin{minipage}{\textwidth}	
	\end{minipage}
 \\[0.5cm]
 \textbf{Study of speaker localization under dynamic and reverberant environments}
	\author[ ]{Daniel A. Mitchell\footnote{mitchdan@post.bgu.ac.il} and Boaz Rafaely\footnote{br@bgu.ac.il}}
  	\affil[ ]{School of Electrical and Computer Engineering, Ben-Gurion University of the Negev, Israel}

}
\date{}
\begin{document}
\clearpage
\setcounter{page}{1}
\maketitle
\thispagestyle{empty}
\fancypagestyle{empty}
{	
	\fancyhf{} \fancyfoot[R]
	{
	}
}
\subsection*{\fontsize{10.5}{19.2}\uppercase{\textbf{Abstract}}}
{\fontsize{10.5}{60}\selectfont Speaker localization in a reverberant environment is a fundamental problem in audio signal processing. Many solutions have been developed to tackle this problem. However, previous algorithms typically assume a stationary environment in which both the microphone array and the sound sources are not moving. With the emergence of wearable microphone arrays, acoustic scenes have become dynamic with moving sources and arrays. This calls for algorithms that perform well in dynamic environments. In this article, we study the performance of a speaker localization algorithm in such an environment. The study is based on the recently published EasyCom speech dataset recorded in reverberant and noisy environments using a wearable array on glasses. Although the localization algorithm performs well in static environments, its performance degraded substantially when used on the EasyCom dataset. The paper presents performance analysis and proposes methods for improvement.}
\noindent{\\ \fontsize{11}{60}\selectfont Keywords: Direction of Arrival (DOA) estimation, Direct Path Dominance (DPD) test, wearable microphone arrays, EasyCom dataset} 

\fontdimen2\font=4pt
\section{\uppercase{Introduction}}\label{sect1}

Localizing multiple sound sources recorded with a microphone array in an enclosure is an important task used in a wide range of applications such as speech enhancement, source separation and video conferencing \cite{brandstein2001microphone}. Therefore, many direction-of-arrival (DOA) estimation methods have been developed for this task. 
These include methods based on beam-forming \cite{van1988beamforming},
 subspace methods such as multiple signal classification (MUSIC) \cite{schmidt1986multiple},
 and time-delay of arrival estimation methods \cite{brandstein1997robust_timedelay}.
 Many of the algorithms based on these methods were designed assuming a free-field environment, but when these algorithms are used in a 
more common reverberant environment their DOA performance degrades. The reason for this is 
 that in a reverberant environment, room reflections mask the direct sound which carries the DOA information.
Recently, however, several methods have been developed for DOA estimation of multiple speakers that are robust to reverberation. 
One such method processes the microphone signals in the time-frequency domain, and employs a direct-path dominance (DPD) test  \cite{madmoni2018improved} to identify time-frequency bins that are dominated by the direct sound from the source. Algorithms which use this method have been widely studied assuming a static environment, where both the sound sources and the microphone array are stationary. On the other hand, these algorithms have been less intensely studied in a more realistic, dynamic environment, where the sound sources and/or the microphone array are moving. 

In dynamic environments, the motion of the sound sources and/or microphone array may lead to the rapid change of DOAs in time. Thus, to accurately trace the DOA of speakers requires a short interval between successive DOA estimates. Additionally, the DOA estimates may be smoothed in time using a tracking algorithm. Although DOA estimation and tracking algorithms in dynamic environments have been the subject of several recent studies \cite{slam2018,wang2013high}, including   the Acoustic Source Localization and Tracking (LOCATA) \cite{lollmann2018locata} challenge, none of these have included experiments with wearable microphone arrays - which may bring new challenges. Such scenes are becoming increasingly popular due to the increased interest in applications involving augmented reality. 

In this article we address the problem of DOA estimation in a noisy dynamic environment involving a wearable microphone array. The experiments were performed using the Easy Communication (EasyCom) dataset \cite{donley2021easycom}, which was explicitly designed to represent a realistic cocktail-party environment. The DOA estimates were computed using a computationally efficient algorithm, which had been shown previously to have good source localization performance in a static reverberant environment \cite{tourbabin2018space}. The algorithm incorporates a DPD test and operates in the time-frequency domain. We study the performance and limitations of this algorithm  on the EasyCom dataset under different operating parameters. We also introduce two modifications of the algorithm and study the consequent improvement in performance.


\section{\uppercase{Mathematical Model}}\label{sect2}
In this section we first briefly present the   model assumed for the recorded signal as captured by the microphone array at each time-frequency bin $(t,f)$. Then,  we describe the local space domain distance (LSDD) algorithm \cite{tourbabin2018space} for the DOA estimation at each bin $(t,f)$. 
\subsection{Signal model}
Assume a microphone array with $M$ microphones arranged according to a pre-defined geometry. Next, consider a sound field comprised of $K$ far-field 
sources, arriving from directions $\Psi_{k}, k \in \{1,2,\ldots,K\}$. These sources represent the direct sound from the speakers in the scene, as well as reflections (reverberations) due to objects and room boundaries. 

In the next step, the recorded microphone signals are transformed into the joint time-frequency domain by applying the short-time Fourier transform (STFT). This is done by first separating the speech signal into short time intervals of length $\delta t$. A fast Fourier transform (FFT) is then applied to each time segment. Following these pre-processing steps, the signal received by the microphone array can then be described in the STFT domain as
\begin{equation}
\textbf{x}(t,f) =  \sum_{k=1}^{K}s_k(t,f)\textbf{v}(f,\Psi_{k})+
\textbf{n}(t,f)\;,
\label{eq:equation1}
\end{equation}
where $\textbf{x}(t,f)=[x_1(t,f), x_2(t,f),\ldots,x_M(t,f)]^{T}$ is an 
$M\times 1$ complex vector denoting  the signal as measured by the microphones composing the array;
$\textbf{v}(f,\Psi_{k})=[v_1(f,\Psi_{k}),v_2(f,\Psi_{k}),\ldots,v_M(f,\Psi_{k})]^{T}$ is an $M\times 1$ vector denoting the response of the microphone array to a unit-amplitude plane wave at frequency $f$ arriving from the $k$th source in direction $\Psi_{k}$; 
$s_k(t,f)$ is a scalar which represents the amplitude of the $k$th sound source signal; and
$\textbf{n}(t,f)=[n_1(t,f), n_2(t,f),\ldots,n_M(t,f)]^{T}$ is an $M\times 1$ vector denoting the noise in the  signal $\textbf{x}(t,f)$.


\subsection{Local Space Domain Distance (LSDD) algorithm}\label{sect2B}
The 
LSDD algorithm 
is a  recently developed DOA estimation algorithm  characterized by DOA performance that is robust to reverberation. The 
algorithm was first proposed in
 \cite{tourbabin2018space} and works as follows. The directional spectrum $ \textbf{S}(t,f)=[S_1(t,f),S_2(t,f),\ldots,S_L(t,f)]^{T}$ computed within this algorithm is an $L\times 1$ vector defined over a grid of DOAs
 $\Theta_{l},l\in \{1,2,\ldots,L\}$. The $l$th component is defined as
\begin{equation}
S_l(t,f)=
 d \left( \textbf{x}\left(t,f\right), \textbf{v}\left(f,\Theta_{l}\right) \right)
=d \left( [x_1(t,f),\ldots,
x_M(t,f)]^{T},[v_1(f,\Theta_{l}),\ldots,v_M(f,\Theta_{l})]^{T} \right)\;,
\label{eq:equation2}
\end{equation}
where $d(\textbf{a},\textbf{b})$ is a function which measures the similarity between two vectors $\textbf{a}$ and $\textbf{b}$. In 
\cite{tourbabin2018space}, $d(\textbf{a},\textbf{b})$ was defined as
\begin{equation}
    d(\textbf{a},\textbf{b})=\frac{1}{{\underset{\beta}\min}\big (\frac{\|\textbf{a}-\beta\textbf{b}\|}{\|\textbf{a}\|}\big)}\;,
    \label{eq:equation3}
\end{equation}
where $\|.\|$ is the 2-norm. However, in this article we shall use 
the more conventional cosine similarity measure:
\begin{equation}
d(\textbf{a},\textbf{b}) = \frac{|<\textbf{a},\textbf{b}>|}{\|\textbf{a}\|\|\textbf{b}\|}\;,
\label{eq:equation4}
\end{equation}
where $<\textbf{a},\textbf{b}>$ denotes the inner product between $\textbf{a}$ and $\textbf{b}$.

\noindent Given the 
spectrum vector $\textbf{S}(t,f)$, 
the estimated DOA for bin $(t,f)$ is computed by
\begin{equation}
\hat{\theta}(t,f) = \underset{l}{\arg \max}\{S_l(t,f)\}\;.
\label{eq:equation5}
\end{equation}
However, some of the bins $(t,f)$ do not contain a valid $\hat{\theta}(t,f)$ value.  These are bins in which the direct signal from the speaker is masked by noise and reverberations. We eliminate these bins by calculating a DPD measure value ${\chi}(t,f)$ for each bin which we then test against a threshold $\lambda$. Although there are different methods available for calculating ${\chi}(t,f)$,  we shall, for simplicity,  use the following:
\begin{equation}
{\chi}(t,f)  = \underset{l}{\max}\left\{S_l(t,f)\right\}\;. 
\label{eq:equation51}
\end{equation}
Together, Eqns. (\ref{eq:equation5}) and (\ref{eq:equation51})   define a (joint) LSDD DOA/DPD algorithm. 
\subsection{Energy Weighted Local Space Domain Distance (LSDDe) algorithm}\label{sect2C}
We describe an energy weighted modification for calculating the DPD test value ${\chi}(t,f)$. In this equation, 
 we weight the  DPD test value in Eqn. (\ref{eq:equation51}) 
with its corresponding signal energy. The energy weighted DPD test value is then:
\begin{equation}
    {\chi}(t,f)=\underset{l}{\max}\left\{S_l(t,f)\cdot MED[|x_1(t,f)|^{2},|x_2(t,f)|^{2},\ldots,|x_M(t,f)|^{2}]\right\}\;, \label{eq:equation55}
\end{equation}
where $MED$ is the median operator. Together, Eqns. (\ref{eq:equation5}) and (\ref{eq:equation55})  define a (joint) LSDDe DOA/DPD algorithm. 
\section{\uppercase{Proposed DOA Estimation Algorithm}}\label{sect3}
We propose a new DOA estimation algorithm which was the outcome of investigating the performance of the LSDD algorithm with the EasyCom dataset. It is clear from Eqns. (\ref{eq:equation2}), (\ref{eq:equation4}) and (\ref{eq:equation5}) that the LSDD algorithm does not use any information about the behavior of
$\textbf{S}(t,f)$  with respect to $\Theta_{l}$. As such information may be useful, it is proposed to incorporate this information using a correlation process,  as follows. For each frequency $f$ we define an ïdeal two-dimensional 
spectrum represented by matrix $\textbf{W}$, whose elements, $W_{lh}\equiv W(\Theta_l,\Theta_h)$, represent the similarity between the $l$th steering vector  $\textbf{v}(f,\Theta_{l})$ and the $h$th steering vector $\textbf{v}(f,\Theta_{h}$) defined as
\begin{equation}
\begin{split}
    W(\Theta_l,\Theta_h) &= d(\textbf{v}(f,\Theta_{l}),\textbf{v}(f,\Theta_{h})) \\
& = d([v_1(f,\Theta_{l}),\ldots,v_M(f,\Theta_{l})]^{T},
[v_1(f,\Theta_{h}),\ldots,v_M(f,\Theta_{h})]^{T})\;, \forall l,h\in\{1,2,\ldots,L\}\;,
\end{split}
    \label{eq:equation6}
\end{equation}
where the measure $d$ is defined in (\ref{eq:equation4}). Now, the similarity between   each column vector in $\textbf{W}$ and the 
spectrum $\textbf{S}(t,f)$  is computed, which provides the following indication.  
Suppose  that  $\Theta_{h}$ is a DOA of an actual source. Then, we expect the $h$th column of 
$\textbf{W}$ to be similar to $\textbf{S}(t,f)$. This, in effect, defines the new directivity based Space Domain Distance (dSDD) DOA estimation algorithm. The corresponding  DOA estimate for bin $(t,f)$ is now computed using
\begin{equation}
\hat{\theta}(t,f) = \underset{h}{\arg \max}\left\{d(\textbf{S}(t,f),\textbf{W}_h)\right\}\;,
\label{eq:equation7}
\end{equation}
where $\textbf{W}_h$ denotes the $h$th column in $\textbf{W}$. We follow Eqn. (\ref{eq:equation51}) and define a corresponding DPD test measure as
\begin{equation}
{\chi}(t,f)  = \underset{h}{\max}\left\{d(\textbf{S}(t,f),\textbf{W}_h)\right\}\;. 
\label{eq:equation71}
\end{equation}
Together,  Eqns. (\ref{eq:equation7}) and (\ref{eq:equation71}) define a (joint)  dSDD DOA/DPD algorithm. It should be noted that under ideal conditions where signal $\textbf{x}$ in Eqn. (\ref{eq:equation1}) is composed of a single plane wave, the two algorithms, LSDD and dSDD, should provide the same estimate as they both rely on the same set of steering vectors. However, the motivation for proposing dSDD is the expected robustness against potential noise and reverberation due to the comparison of entire functions, or vectors. This is in contrast to the LSDD where DOA estimates are based on looking for a peak in a function. \newline 

\noindent As in the case of the LSDD algorithm, we describe an energy weighted dSDD algorithm (dSDDe),
 in which, for each  bin $(t,f)$, we weight the  dSDD DPD test measure 
with the corresponding signal energy. The energy weighted DPD test value is therefore given by
\begin{equation}
{\chi}(t,f)=\underset{h}{\max}\left\{d(\textbf{S}(t,f),\textbf{W}_h)\right\}\cdot MED\left[ |x_1(t,f)|^{2},|x_2(t,f)|^{2},\ldots,|x_M(t,f)|^{2} \right]\;.
\label{eq:equation75}
\end{equation}
\section{\uppercase {Experiments}}\label{sect4}  
This section presents an experimental study that aimed to investigate the performance of the LSDD and the dSDD algorithms with the EasyComm dataset. First, the experimental setup and methodology are presented. This is then followed by the evaluation of the results. 
\subsection{Set-up}\label{Sect41} 
\begin{figure}[H]
    \centering
    \includegraphics[scale=0.06]{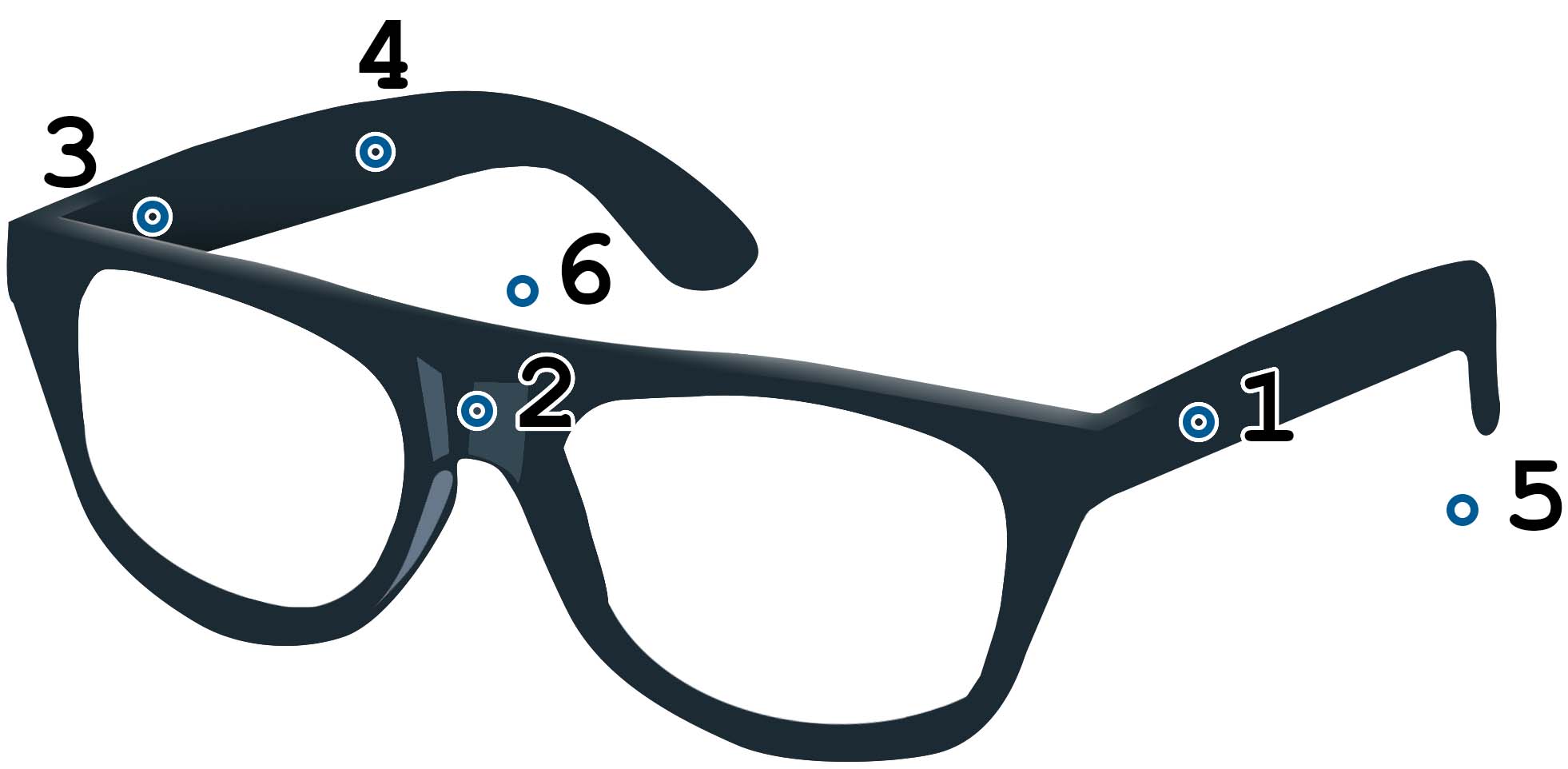}
    \caption{Illustration of the AR glasses with locations of microphones  \cite{donley2021easycom}.  Four of the microphones are fixed rigidly to the glasses and two of the microphones are placed in the user's ears.}
    \label{fig:glasses}
\end{figure}

The experiments described in this article were performed on the 
EasyCom dataset  
\cite{donley2021easycom}. This 
dataset was designed with the aim of analyzing the cocktail party effect with audio signals captured by augmented reality (AR) glasses equipped with an egocentric six-channel microphone array. Figure \ref{fig:glasses} shows a schematic drawing of the glasses with locations of the microphones \cite{donley2021easycom}.

The dataset contains recordings of natural conversations in a noisy restaurant environment. Participants were equipped with close-talk microphones, a camera and tracking markers. They were asked to engage in conversations during several tasks, including introductions, ordering food, solving puzzles, playing games and reading sentences. The recordings also contain an egocentric video viewpoint of the participants. The pose (position and rotation) of every participant was also recorded. The dataset was additionally labelled with annotators of voice activity.
The signals recorded by the microphones were sampled at a rate of $48 \textrm{ kHz}$. The recorded data was transformed into the STFT domain using a 1024 samples ($\simeq 20 \textrm{ msec}$) Hann window with an overlap of 512 samples. The microphone signals in the STFT domain were employed as an input to the algorithms under study. 

\subsection{Methodology}
Evaluation of the DOA/DPD algorithms incorporated a direction search with a resolution of $6^\circ$, which was limited to the horizontal plane. The ground truth azimuthal DOA ($\Psi_{k}$) was obtained from the EasyCom dataset as a function of time. Altogether,  a series of three experiments was performed with the EasyCom dataset. The first experiment measured the effective frequency range  of the array $[f_{low},f_{high}]$, and the second and third experiments investigated the effect of frequency smoothing and the length of the time interval $\Delta T$ on the performance.

DOA estimation performance was evaluated as follows. For each $(t,f)$ bin, the absolute error, 
\begin{equation}
    \varepsilon(t,f)=|\Psi(t) - \hat{\theta}(t,f)|\;,
    \label{eq:error}
\end{equation}
between the true DOA $\Psi(t)$  and the estimated DOA $\hat{\theta}(t,f)$ was computed. Note that Eqn. (\ref{eq:error}) assumes that both $\Psi(t)$ and $\hat{\theta}(t,f)$ are measured with respect to the same axis. In practice, in this dataset,  $\Psi(t)$ is measured with respect to an  axis defined relative to the room, while $\hat{\theta}(t,f)$ is measured with respect to the orientation of the glasses. Thus, before calculating $\varepsilon(t,f)$, $\hat{\theta}(t,f)$ is transformed to the fixed axis of the room by incorporating head tracking information. In addition, each bin was labeled a "hit" if $\varepsilon(t,f)\leq 10^{\circ}$; otherwise it was labeled a "miss".

The bins ($t,f$) were divided into time blocks of length $\Delta T$. For each block, only bins which satisfied the following four conditions were regarded as valid and were used for DOA estimation:
\begin{enumerate}
    \item The frequency $f$ lies within the effective operating frequency range: $f\in [f_{low},f_{high}]$.
    \item The time $t$ lies within the selected block. Denoting $T$ as the middle time of the block, then $t\in [T-\Delta T/2, T+\Delta T/2]$ belongs to block $T$.
    \item Voice activity was detected at time $t$.
    \item The DPD test value $\chi(t,f)$ exceeds a threshold $\lambda$. For the purpose of this study, the threshold was determined from the percentage $p$ of bins which satisfy the first three conditions with the highest value $\chi$. Note that $
    \lambda$ is computed independently for each block $T$.
\end{enumerate}
The mean value of $\varepsilon(t,f)$ was computed over valid $(t,f)$ bins, i.e. bins which satisfy the above four conditions. By definition, this is the mean absolute DOA estimation error $E(p,T,\Delta T)$ for the block $T$. Similarly, the mean hit ratio $H(p,T,\Delta T)$ was computed on valid $(t,f)$ bins by dividing the number of valid bins labelled "hit"  by the total number of valid bins. Finally, the  mean absolute error, $\bar{E}(p,\Delta T)$, and the mean hit ratio, $\bar{H}(p,\Delta T)$, for the entire experiment was computed by averaging $E(p,T,\Delta T)$ and $H(p,T,\Delta T)$ over all  blocks $T$.  

\subsection{Effective operating frequency band }
\begin{figure}[H]
    \centering
    \includegraphics[scale=0.4]{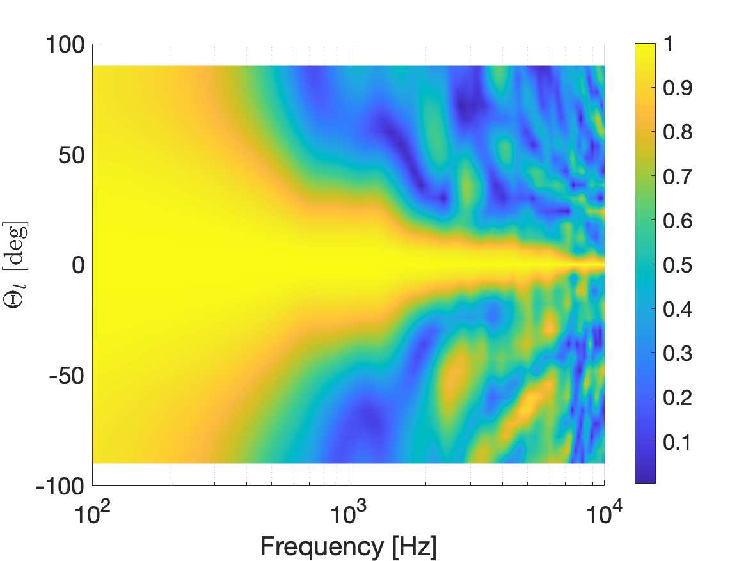}
    \caption{The similarity matrix $\Lambda$ for $\Theta_{h} = 0^\circ$.}
    \label{fig:beampatten}
\end{figure}

The EasyCom dataset involves speech sound which naturally limits the frequency range of interest  \cite{monson2014perceptual}. This frequency band is reduced in practice by aliasing effects which arise
from the microphone array.
For a specific steering vector $\textbf{v}(f,\Theta_{h})$ (corresponding to a frequency $f$ and direction $\Theta_{h}$), the similarity between $\textbf{v}(f,\Theta_{h})$ and the set of steering vectors $\textbf{v}(f,\Theta_{l}), l \in \{1,2,...,L\}$ was computed using Eqn. (\ref{eq:equation4}). This was repeated for all frequencies $f$ leading to the following measure:
\begin{equation}
    {\Lambda}(f,\Theta_l) = d(\textbf{v}(f,\Theta_{h}),\textbf{v}(f,\Theta_{l}))\;.
\end{equation}
Figure \ref{fig:beampatten} shows $\Lambda$ for $\Theta_{h}=0^{\circ}$. Visual inspection shows that the preferred frequency band is about $1100 - 2000$ Hz, where at lower frequencies the  directivity may be too wide, while at higher frequencies significant side lobes may degrade spatial processing. While this is a relatively narrow band of frequencies, in this work it led to the best performance. Extending the range of operation for both lower and higher frequencies is proposed for future work.
\subsection{DOA error and frequency smoothing} \label{sec:freqSmooth}
The spectrum  $\textbf{S}(t,f)$  plays a key role in the DOA/DPD algorithms. In particular, the authors in \cite{beit2020importance} have shown that it is beneficial  to smooth 
$\textbf{S}(t,f)$ in frequency.
We investigated the effect of smoothing $\textbf{S}(t,f)$ over frequency using a moving average filter of length $(2R+1)$. Let 
$\bar{\textbf{S}}(t,f)=
[\bar{S}_1(t,f),\bar{S}_2(t,f),\ldots,\bar{S}_L(t,f)]^{T}$ denote the smoothed spectrum, which is computed by
\begin{equation}
    \bar{S}_l(t,f)=\sum_{r=-R}^{R}S_l(t,f+r\Delta f)/(2R+1)\;,
\end{equation}
where $\Delta f$ is the STFT frequency bin width and $\textbf{S}_l(t,f)$ is defined in Eqn.  (\ref{eq:equation2}).

\begin{figure}[H]

     \centering
     \begin{subfigure}[b]{0.3\textwidth}
         \centering
         \includegraphics[width=\textwidth]{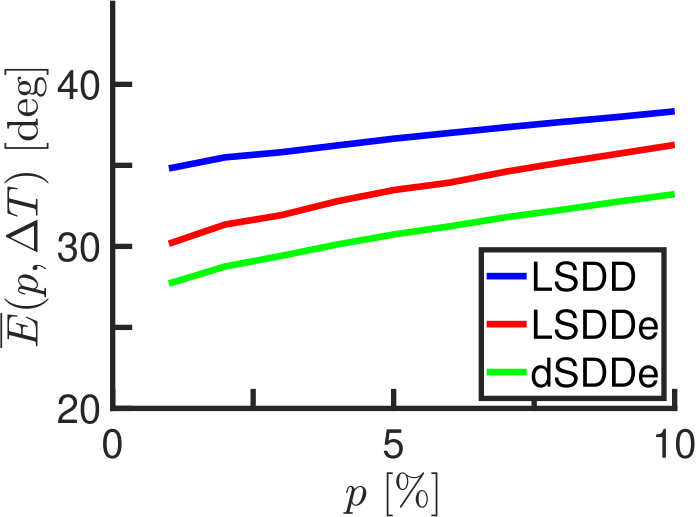}
         \caption{ }
         \label{fig:five sin x}
     \end{subfigure}
     \hfill
     \begin{subfigure}[b]{0.3\textwidth}
         \centering
         \includegraphics[width=\textwidth]{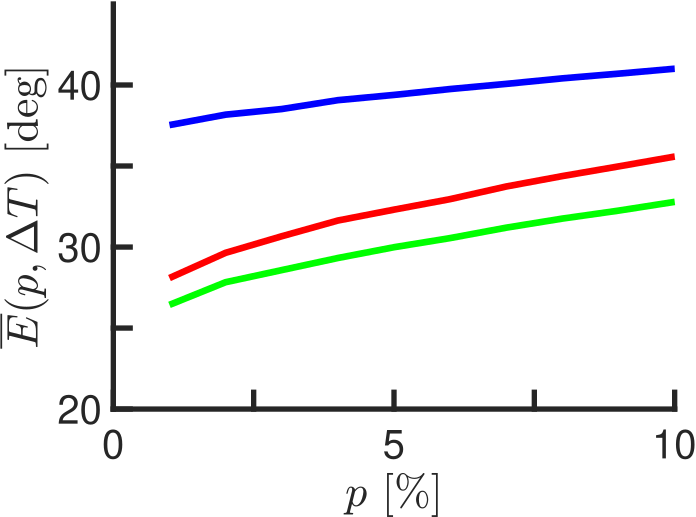}
         \caption{ }
         \label{fig:five over x}
     \end{subfigure}
     \hfill
     \begin{subfigure}[b]{0.3\textwidth}
         \centering
         \includegraphics[width=\textwidth]{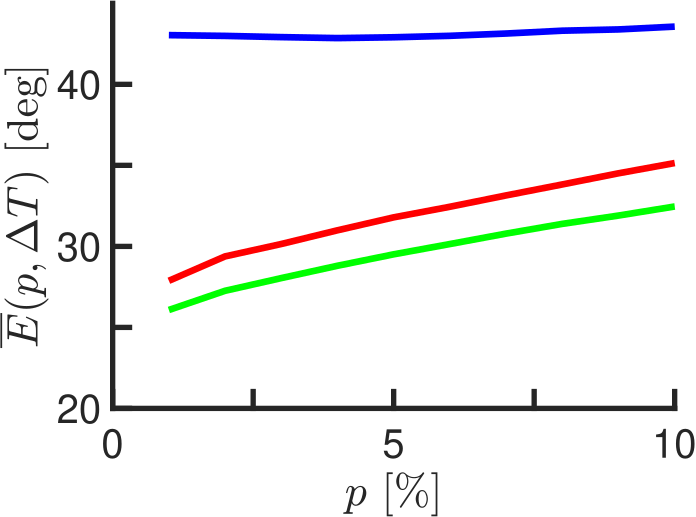}
         \caption{ }
         \label{fig:five over x}
     \end{subfigure}
         \begin{subfigure}[b]{0.3\textwidth}
         \centering
         \includegraphics[width=\textwidth]{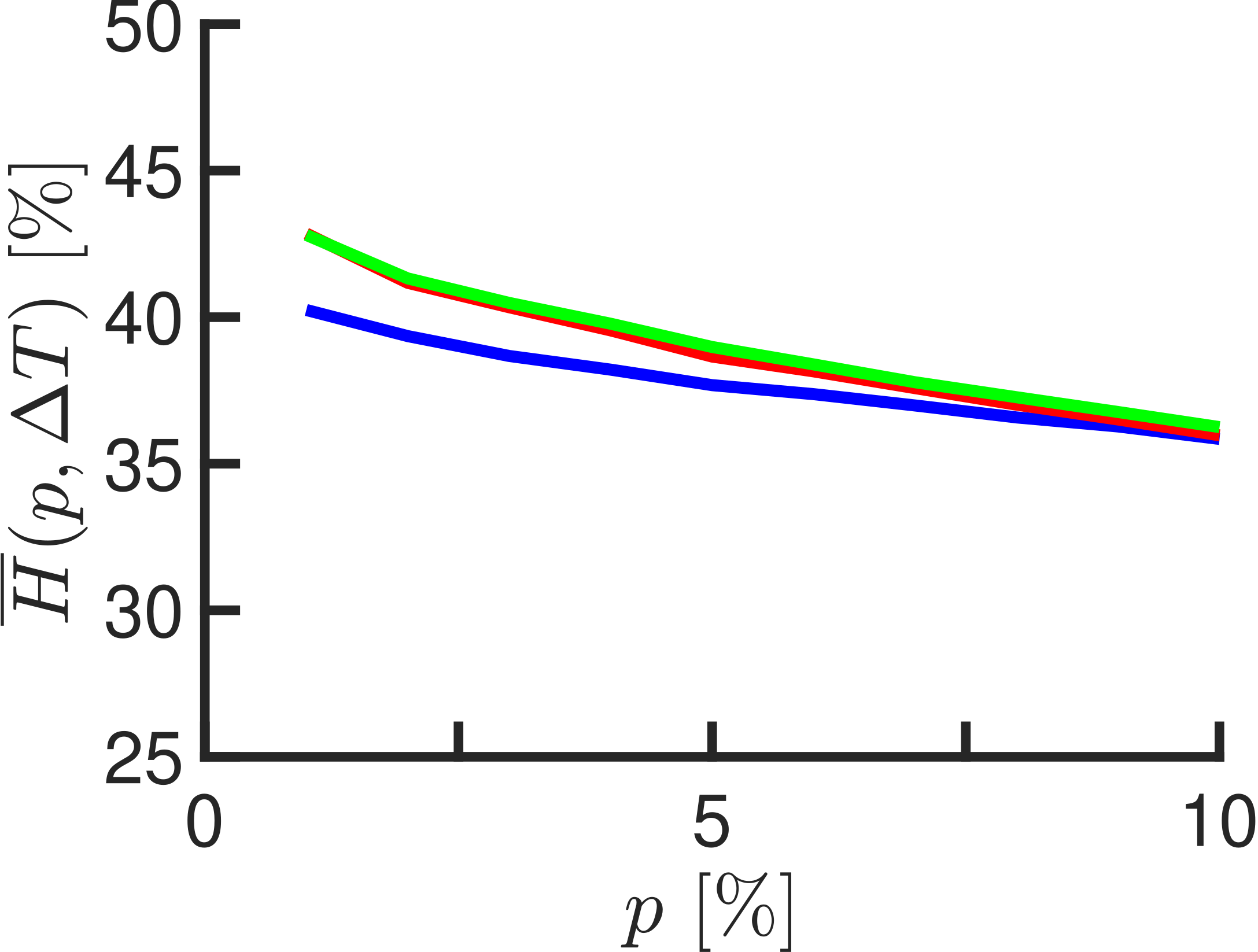}
         \caption{ }
         \label{fig:five sin x}
     \end{subfigure}
     \hfill
     \begin{subfigure}[b]{0.3\textwidth}
         \centering
         \includegraphics[width=\textwidth]{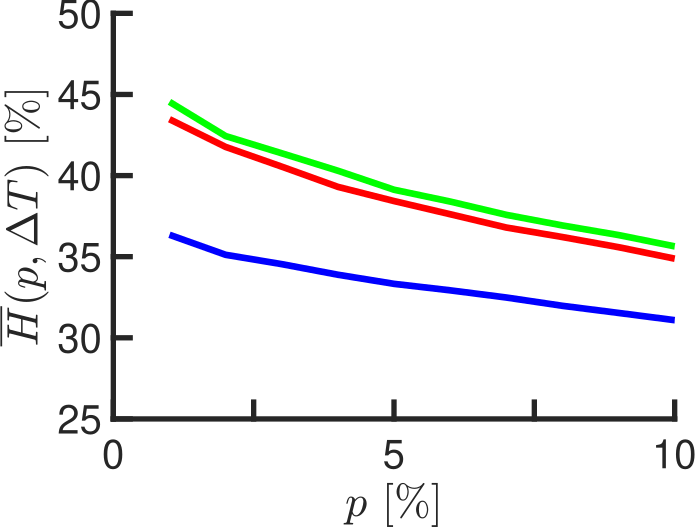}
         \caption{ }
         \label{fig:five over x}
     \end{subfigure}
     \hfill
     \begin{subfigure}[b]{0.3\textwidth}
         \centering
         \includegraphics[width=\textwidth]{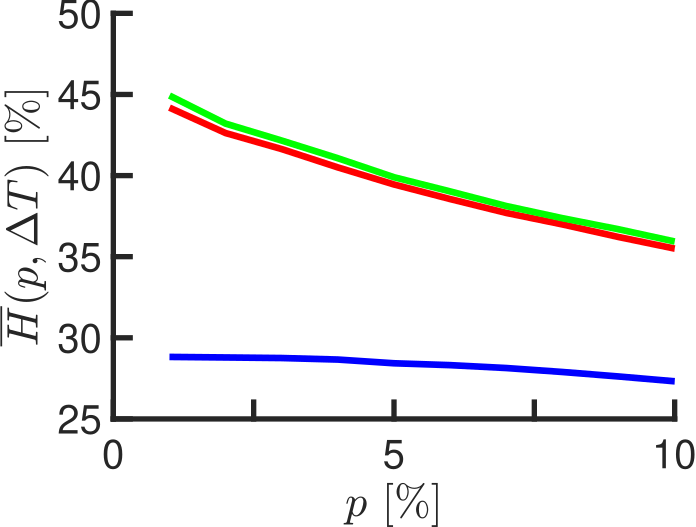}
         \caption{ }
         \label{fig:five over x}
     \end{subfigure}
        \caption{The mean absolute error $\bar{E}(p,\Delta T)$ and the mean hit ratio $\bar{H}(p,\Delta T)$  as a function of  $p$ 
        when averaged over ten 1-minute data segments. The time interval was $\Delta T = 200$ msec. (a)-(c)  Show, respectively, $\bar{E}(p, \Delta T)$   with a $9$-element smoothing filter, a $3$-element smoothing filter and no smoothing. (d)-(f) show the corresponding $\bar{H}(p,\Delta T)$  curves.}
        \label{fig3}
\end{figure}

DOA estimation experiments were carried out on several 1-minute segments extracted from the EasyCom dataset. Altogether, $\textbf{S}(t,f)$ was frequency smoothed using (i) a 9-element filter, (ii) a 3-element filter and (iii) no smoothing. In this experiment,  $\Delta T$ was fixed to $200$ msec. The corresponding results are shown in  Fig. \ref{fig3}. 
Figure 3 shows that best overall performance, in terms of both $\bar{E}(p,\Delta T)$ and $\bar{H}(p,\Delta T)$, for the original LSDD algorithm as described in Sec. \ref{sect2B} is obtained with the $9$-element filter.  On the other hand,  for  both LSDDe as in Sect. \ref{sect2C} and  dSDDe as in Sect. \ref{sect3}, the best overall performance is obtained when no smoothing is used. Moreover, for both algorithms the change in performance with respect to smoothing is relatively small. 

Finally, we compare the results obtained with the variation of the LSDD algorithm (i.e with a $9$-element smoothing filter) and the best variations of the LSDDe and dSDDe algorithms (i.e. no smoothing). At low percentages $p$ the mean absolute error $\bar{E}(p,\Delta T)$ obtained with the dSDDe algorithm is approximately $9^{\circ}$ lower than that obtained with the LSDD algorithm. Similarly, the mean   hit ratio obtained with the dSDDe algorithm is approximately $5 \%$  higher than that obtained with the LSDD algorithm.

\subsection{DOA error and time interval $\Delta T$}
\begin{figure}[H]
     \centering
     \begin{subfigure}[b]{0.3\textwidth}
         \centering
         \includegraphics[width=\textwidth]{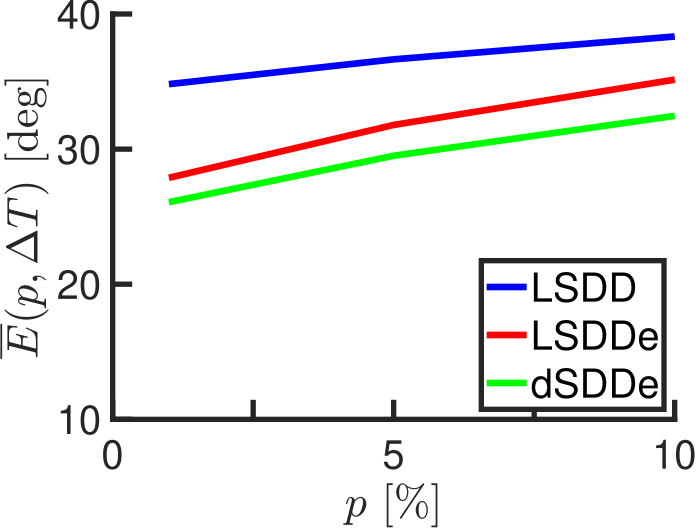}
         \caption{ }
         \label{fig:five sin x}
     \end{subfigure}
     \hfill
     \begin{subfigure}[b]{0.3\textwidth}
         \centering
         \includegraphics[width=\textwidth]{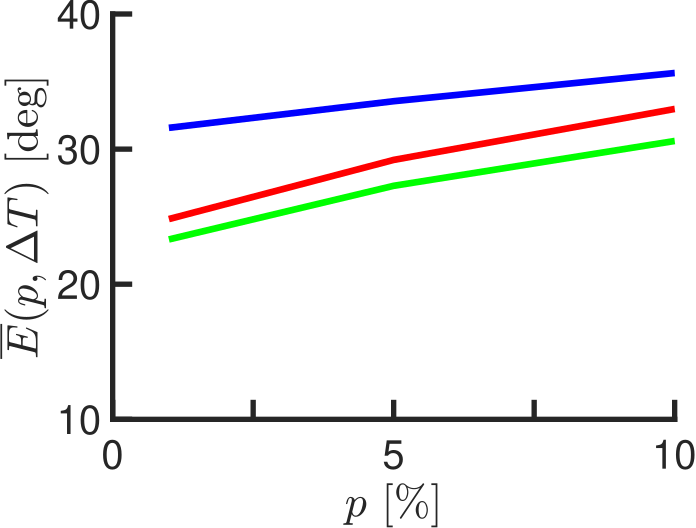}
         \caption{ }
         \label{fig:five over x}
     \end{subfigure}
     \hfill
     \begin{subfigure}[b]{0.3\textwidth}
         \centering
         \includegraphics[width=\textwidth]{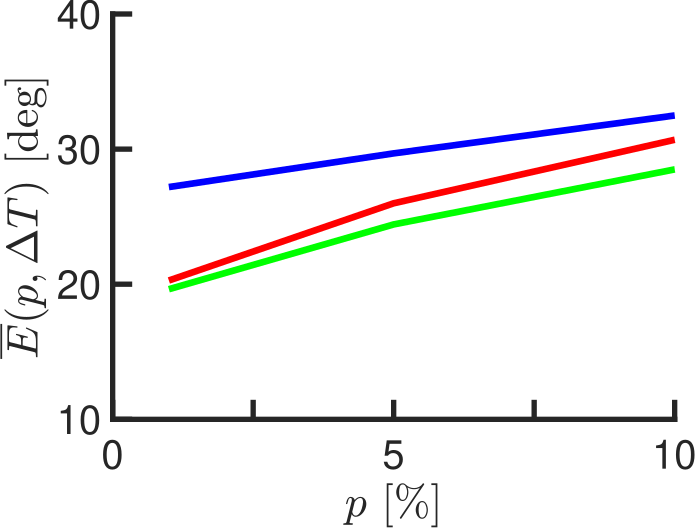}
         \caption{ }
         \label{fig:five over x}
     \end{subfigure}
        \caption{ $\bar{E}(p,{\Delta T})$ as a function of percentage $p$ for the LSDD, LSDDe and dSDDe algorithms for three different 
        ${\Delta T}$ values, averaged over ten 1-minute segments. (a) ${\Delta T} = 200$ msec; (b) ${\Delta T} = 300$ msec; (c)  ${\Delta T} = 500$ msec. Best  frequency smoothing was applied as in Sec. \ref{sec:freqSmooth}.
        } 
        \label{fig4}
        \end{figure}

The choice of time interval $\Delta T$ may be directly related to the dynamic nature of the dataset. In general, we would like to use a value of $\Delta T$ which is small enough such that the environment can be considered spatially stationary within the interval. However, if the chosen value of $\Delta T$ is too small, DOA performance may degrade. In this experiment, three different values of  $\Delta T$ were investigated:  (a)  $\Delta T = 200$ msec; (b) $\Delta T = 300$ msec; and (c) $\Delta T = 500$ msec. The DOA/DPD algorithms used the best frequency smoothing as presented in the previous section. Figure \ref{fig4} shows $\bar{E}(p,{\Delta T})$ averaged over ten 1-minute segments for three $\Delta T$ values. The figure shows  a consistent and significant improvement in performance as   $\Delta T$ increases.  At $\Delta T=500$ msec and $p=1\%$, the dSDDe algorithm gave a mean absolute DOA error $\bar{E}\approx 20^{\circ}$. Similar to Fig. 3, the dSDDe algorithm achieved the best  performance, and the LSDD algorithm the worst.  

\subsection{Summary}
\begin{figure}[H]
    \centering
    \includegraphics[scale = 0.6]{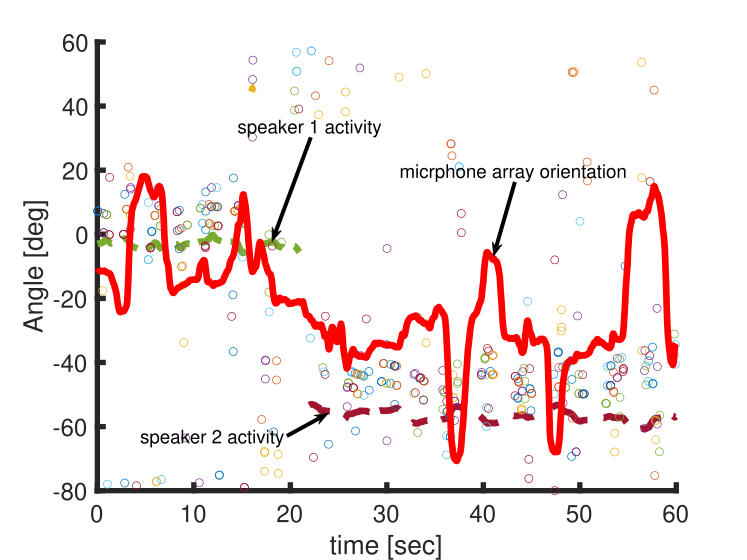}
    \caption{Ground-truth and estimated DOAs for a 1-minute long segment taken from the EasyCom dataset, including two speakers. Ground-truth directions are presented using dashed thick lines. The orientation of the microphone array is marked using a thick red line. The DOA estimates $\hat{\theta}(t,f)$ provided by the dSDDe algorithm with $p=1\%$ and $\Delta T = 200$ msec   are shown as colored circles. 
    }
    \label{fig:dynamic}
\end{figure}

Figure \ref{fig:dynamic} illustrates the performance of the dSDDe algorithm together with a timeline of several experimental data values, for an example taken from the EasyCom dataset.  In this example there are
two active speakers. The time interval 
$\Delta T$ used in the figure is $200$ msec.  
The figure illustrates the dynamic nature of the environment. While many of the DOA estimates fall close to active speakers, some fall very far away from the true DOA. There is also a clear bias in the estimation toward a preferred direction. These findings raise the need for further research to better understand the dataset and the algorithms, and propose improved solutions.  
\section{\uppercase {Conclusions}}\label{sect6}
This work presented three experiments for DOA estimation based on the EasyCom dataset. These preliminary experiments showed that: 
\begin{enumerate}
\item The baseline performance using the original LSDD algorithm showed limited performance.
\item The dSDD algorithm improved performance by incorporating more detailed spatial information.
\item Energy weighting DPD test values was found to be useful.  
\item Both the LSDDe and the dSDDe DOA algorithms do not require frequency smoothing.
\end{enumerate}
\renewcommand{\refname}{\normalfont\selectfont\normalsize}

\section*{\uppercase {Acknowledgements}}\label{sect6}
This research was supported by Reality Labs at Meta.
\noindent \section*{\uppercase{References}}
\vspace{-18pt}
\bibliographystyle{abbrv}
\bibliography{mybibliography}

\end{document}